\documentstyle[pra,aps]{revtex}

\newcommand{\be}{\begin{equation}}
\newcommand{\ee}{\end{equation}}
\newcommand{\br}{\begin{eqnarray}}
\newcommand{\er}{\end{eqnarray}}
\newcommand{\nn}{\nonumber}
\newcommand{\bd}{\begin{displaymath}}
\newcommand{\ed}{\end{displaymath}}

\newcommand{\bfig}{\begin{figure}}
\newcommand{\efig}{\end{figure}}
\def\alf{\alpha}

\def\lb#1{\label{#1}}
\def\3cdot{\cdot \cdot \cdot}
\def\rf#1{\ref{#1}}

\def\d#1{\delta_{#1}}

\def\l{\lambda}

\def\o{\omega}
\def\O{\Omega}
\def\om0{\omega _0}
\def\Om0{\Omega _0}

\def\rg{\rangle}
\def\lg{\langle}
\def\text#1{{\rm{#1}}}

\def\d{\delta}

\def\->{\rightarrow}
\def\=>{\Rightarrow}
\def\-->{\longrightarrow}
\def\==>{\Longrightarrow}

\def\rpar{\right)}
\def\lpar{\left(}
\def\lbk{\left[}
\def\rbk{\right]}
\def\lbr{\left\{}
\def\rbr{\right\}}
\def\dag{\dagger}
\def\ox{\otimes}

\def\til{\tilde}

\def\pr{^\prime}
\def\pr2{^{\prime\prime}}
\def\rf#1{(\ref{#1})}
\def\Re{{\rm Re}}
\def\Im{{\rm Im}}

\def\bfig{\begin{figure}}
\def\efig{\end{figure}}
\def\jour#1#2#3#4{{#1} {\bf #2}, #3 (#4)}
\def\AJP{\em Am. J. Phys. }
\def\AP{\em Ann. Phys. (N.Y.)}

\def\NCB{{\em Nuovo Cimento} B}

\def\PLA{{\em Phys. Lett.}  A}
\def\PRL{\em Phys. Rev. Lett.}
\def\PR{\em Phys. Rev.}
\def\PRA{{\em Phys. Rev.} A}

\def\PRD{{\em Phys. Rev.} D}

\def\PA{{\em Physica} A}
\def\PB{{\em Physica} B}
\def\PT{\em Physics Today}

\def\JPA{\em J. Phys. A: Math. Gen.}

\begin{document}
\draft
\title{Information Transfer in the Course of a Quantum Interaction}
\author{Marcos C. de Oliveira$^1$\thanks{E-mail: marcos@physics.uq.edu.au},
        Salomon S. Mizrahi$^2$\thanks{E-mail: salomon@power.ufscar.br},
        Victor V. Dodonov$^2$\thanks{E-mail: vdodonov@power.ufscar.br}}
\address{$^1$Centre for Laser Science, Department of Physics, \\
University of Queensland, QLD 4072, Brisbane, Australia. \\
$^2$Departamento de F\'\i sica, CCT, Universidade Federal de S\~ao Carlos,\\
Rod. Washington Luiz Km 235, S\~ao Carlos, 13565-905, SP, Brazil.\\
}
\date{\today}

\maketitle

\begin{abstract}
We discuss the problem of the information transfer (exchange of states
configuration) between two interacting quantum systems along their
evolution in time. We consider the specific case of two modes of the
electromagnetic field with rotating wave coupling interaction
(up-conversion in a nonlinear crystal). We verify that for certain
initial states of the fields the swapping of state configuration
occurs with conservation of the mean energy of each mode
(without energy transfer), characterising thus pure information flow.
 
\end{abstract}



%

\vspace{5mm}

\section{introduction}
%
The quantum two-mode harmonic oscillator (HO) allies simplicity together
with richness of details and more importantly it is associated
with a large variety of physical phenomena. Many properties of
this system were studied in detail in the classical papers on quantum
amplifiers and converters
\cite{Louis1,Gord,mollow1,mollow2,mollow3,3walls,Lu}.
Among later publications we may refer to the considerations of
various time-dependent couplings in
\cite{glauber1,Abd1,Sand,Yeon,Abd2,Kimman,Oman,Lo,dmm95,Abd3}.
The interaction between two electromagnetic modes, each one in a cavity,
was considered in the analysis of reversible decoherence of superposition
of field states in \cite{raimond}.
Quantum-statistical properties of two coupled modes of electromagnetic
field were studied recently in \cite{abd4,Veis,Kal}.
A simple model of two coupled oscillators was used in \cite{Han} to
illustrate Feynman's concept of the ``rest of the universe''.
In quantum optics, one encounters the phenomena of two electromagnetic
(EM) fields interacting in a nonlinear crystal \cite{yariv,3louis,perina},
or through a beam splitter, or yet a
coupler \cite{lai}. The two-mode HO also apppears in problems
involving an ion trapped in a two-dimensional well, with
the ion center of mass oscillating harmonically and interacting with
two electronic levels \cite{steinbach}.

In another scenario, the problem of two-level atoms (the ``reservoir")
interacting {\em dispersively} with one EM mode in a cavity was
discussed in \cite{brune1}.
In their approach the field state is projected on photon number eigenstate
without energy exchange between atom and field. The change in the
field photon probability distribution is achieved through a dissipation
free ``information-gathering" process.

However, for our purpose of approaching pure information transfer
between quantum subsystems, we are adopting here quantum optics
(two EM fields at different frequencies and a classical driving field
interact) as the scenario where the physical phenomena take place.
We analyse the dynamical evolution of a two-mode
harmonic system (interacting {\em resonantly} with the driving field)
for some particular initial joint-state, one mode being a
HO coherent state and the other is a special superposition
of two coherent states, known as `cat' state \cite{buzek}. We are
interested on two aspects of the quantum evolution, i) the
{\em recurrence} to the initial state and ii) how and when
information is transfered from one mode to the other, such that the
states of the two modes exchange identity (or states configuration),
meaning that during their evolution, at some times $\tau _n$, they
assume each other initial state, thus transfering completely their
characteristics to one another. Although the interacting two-mode
problem was already approached in the literature, we did not see it
discussed as in the present physical context;
in a more restricted sense a similar problem was considered earlier in
\cite{3walls,Lu,Abd3}, however, the authors of those papers were
interested only in the energy exchange, and did not
discuss the exchange of the quantum states configuration of the modes,
as a pure transference of information. We will show that when both modes
have the same number of quanta (not necessarily with the same energy)
the information flows in two directions, without expense of energy
for the modes. However, the variances of the number of photons oscillate
in time, being a consequence of the exchange of the
characteristics of the modes. Moreover, if the two fields have the
same phase, looking at the reduced density matrix of one mode (call
it $A$), we verify that the linear entropy functional of $A$ is
independent of mode $B$, this tells that when the mean energy of
each oscillator remains constant during the evolution,
the functional describes exclusively the flow of information to and from
$A$. We also introduce a functional that measures the `degree of
{\em exchangeness}' between the modes.

The paper is planned as follows. In section II we consider the
{\em rotating wave coupling} between two modes of the EM field in a
nonlinear crystal and solve the dynamical equations for the mode
operators. In section III we derive the joint-state vector and write
it down at some characteristic times for the sake
of comparison. In section IV we discuss the decoherence and entropy
for one mode, the other mode assuming the role of `environment'.
In section V we discuss the intermode exchange of information with and
without energy tranfer and section VI contains a summary and our
conclusions.

\section{Dynamics of the two-mode system and characteristic times}
%
In a good approximation, the hamiltonian describing the interaction
between two modes of EM field in the presence of a strong classical
pumping in a nonlinear crystal can be written as
\cite{Louis1,yariv,3louis,perina}
\begin{eqnarray}
H &=&\hbar \omega _{a}a^{\dagger }a+\hbar \omega _{b}b^{\dagger }b+2\hbar
\lambda \cos (\nu t-\phi )\left( a^{\dagger }+a\right) \left( b^{\dagger
}+b\right)  \nonumber \\
&=&\hbar \omega _{a}a^{\dagger }a+\hbar \omega _{b}b^{\dagger }b+\hbar
\lambda \left[ e^{i\left( \nu t-\phi \right) }\left(
a^{\dagger }b^{\dagger }+ab+a^{\dagger }b+ab^{\dagger }\right)  
+ h.c. \right] \nonumber\\
&\equiv& H_a +H_b +V_I
\, ,  \lb{H}
\end{eqnarray}
where h.c. means hermitean conjugate, the coupling parameter $\l$ is
the product of the pump amplitude and the coupling constant between
the EM-field and the crystal, and $\nu$ is the pump field
frequency. The operators $a$, $b$ and
$a^\dag$, $b^\dag$ are the destruction and creation of the field quanta.

Since the Hamiltonian (\ref{H}) is quadratic, the solutions of the
Heisenberg equations of motion, $a(t)$ and $b(t)$, are linear combinations
of {\it all four} initial operators, $a(0)$, $b(0)$,
$a^\dag(0)$, $b^\dag(0)$, with time-dependent coefficients \cite{mollow3}.
There are, however, two important special cases, when these linear
combinations have {\it only two} terms.

If $\nu \approx\omega _{a}+\omega _{b}$, then only the terms
proportional to the products $e^{i\nu t}ab$ and $e^{-i\nu t}a^\dag b^\dag$
in the interaction part of the Hamiltonian (\ref{H})
contribute effectively to the process (the free operators
evolve in the Heisenberg picture as $a_0(t)=e^{-i\o_a t}a(0)$
and $b_0(t)=e^{-i\o_a t}b(0)$). Thus omitting all non-essential
terms one arrives at the effective
{\em counterrotating wave coupling\/} (CWC),
\begin{equation}
V_{CWC}= \hbar \lambda \left[ e^{i(\phi - \nu t)}
a^{\dagger }b^{\dagger }+e^{i(\nu t-\phi) }ab\right] \, , \lb{cgirante}
\end{equation}
which describes the parametric amplification
\cite{Louis1,Gord,mollow1,mollow2} or parametric down conversion.
In this case, $a(t)$ is a linear superposition of $a(0)$ and $b^\dag(0)$
only, the coefficients being {\it hyperbolic\/} functions of time
(if $\nu =\omega _{a}+\omega _{b}$), so no exchange between
operators $a$ and $b$ is possible.

We concentrate on another special case,
$\nu \approx \omega _{a}-\omega _{b}$,
which corresponds to the {\it up-conversion\/} process.
In this case, the most significant part of the interaction Hamiltonian
is given by the {\it rotating wave coupling\/} (RWC)
\begin{equation}
V_{RWC} =\hbar \lambda \left[ e^{i(\nu t-\phi)}
ab^{\dagger }+e^{i(\phi -\nu t)}a^{\dagger }b\right] \, , \lb{girante}
\end{equation}
so the equations of motion for the operators $\tilde{a}(t)=a(t)e^{i\o_a t}$
and $\tilde{b}(t)=b(t)e^{i\o_b t}$
(which correspond to the interaction picture) read
\br
\frac{d\,\tilde{a}}{dt} &=&-i\lambda
e^{i(\Omega t +\phi) }\tilde{b} \,,  \label{dera}
\\
\frac{d\,\tilde{b}}{dt} &=&-i\lambda
e^{-i(\Omega t +\phi) }\tilde{a},  \label{derb}
\end{eqnarray}
where $\O \equiv \o _a - \o _b - \nu \ll \o_a , \o_b$.
Recently, Senitzky \cite{senitzky} analysed the
physical nuances of the zero-point energy, considering four
different kinds of two-mode interaction, identifying the presence
of the van der Waals attraction in all but the rotating wave-coupling
interaction. Nevertheless it is exactly this coupling that will play
an essential role here, becoming the focus of our attention.

Deriving equation \rf{dera} with respect to time and substituting then
equation \rf{derb},
one obtains a second order differential equation to $\tilde{a}$ only,
\begin{equation}
\frac{d^{2}\tilde{a}}{dt^{2}}-i\Omega \frac{d\,\tilde{a}}{dt}+
\lambda ^{2}\tilde{a}=0.  \label{sola}
\end{equation}
Solving (\ref{sola}) with the account of the initial conditions
and returning back to the Heisenberg operators $a(t)$ and $b(t)$,
we obtain
\begin{eqnarray}
a(t;\tau) &=& e^{-i\o_a t} \left[u_{1}(\tau)a(0)+v_{1}(\tau)b(0)\right]
= e^{-i\o_a t} \til{a}(\tau)\, ,
\label{at} \\
b(t;\tau) &=& e^{-i\o_b t} \left[u_{2}(\tau)b(0)+v_{2}(\tau)a(0)\right]
=e^{-i\o_b t}\til{b}(\tau) \, ,
\label{bt}
\end{eqnarray}
where the time-dependent coefficients are given by
\begin{equation}
u_{1}(\tau)=e^{i\chi \tau}\lbk \cos \tau -i \chi \sin \tau \right]\, ,
\qquad u_{2}(\tau)= u_1^{*}(\tau)\, ;  \label{u}
\end{equation}
\begin{equation}
v_{1}(\tau)=-i\sqrt{1-\chi^2} e^{i(\chi \tau +\phi) }
\sin  \tau \, , \qquad v_{2}(\tau)= - v_1^{*} (\tau) \,,  \label{v}
\end{equation}
and
\be
\tau \equiv \o t ~, \quad \chi \equiv \frac{\O}{2 \o} \quad 
\quad \o \equiv \sqrt{\O ^2 + 4\l ^2} /2 \, .
\label{deftau}
\ee
Now we notice that the RWC approximation is justified provided the coupling
and detuning are much smaller than the fields frequencies, i.e.
$\l,\O\ll \o_a,\o_b$. Under these
conditions, the true time $t$ and the dimensionless `slow time' $\tau$
can be considered actually as independent variables, since even a shift of
$t$ by $\pi/\o_{a,b}$ practically does not affects the value of $\tau$ and,
consequently, the values of coefficients $u_j(\tau)$ and $v_j(\tau)$,
$j=1,2$.

Looking at equation (\ref{v}) we notice that for the values of `slow time'
$\tau_n=n\pi,~n=1,2,3,...$ the coefficients $v_{1,2}(\tau)$ turn into
zero, so the operators $a(t)$ and $b(t)$ assume their initial values,
up to unitary phase factors:
\br
\til{a}(\tau_n) &=&a(0)\exp\left[ in\pi(1+\chi) \right],
\label{recora}\\
\til{b}(\tau_n) &=&b(0)\exp\left[in\pi(1-\chi) \right].
\lb{recorb}
\er
These results are in agreement with the quantum recurrence theorem as
proposed by Bocchieri and Loinger \cite{bocchieri} and
$T =\pi / \o $ is the {\em recurrence time}.

Another important characteristic time of the two-mode system is the
period when an exchange (but for a phase factor) of the two modes operators
takes place, namely,
\be
\tilde{a}(\tau_n^{\prime }) = e^{i\theta _n }b(0) \, , \quad
\tilde{b}(\tau_n^{\prime }) = e^{i\d _n }a(0) \, .
\ee
These conditions imply $u_1(\tau_n^{\prime })=u_2(\tau_n^{\prime })=0$,
which can happen only provided $\O=\chi=0$ and 
$\tau_n^{\prime }=(n-1/2)\pi,~n=1,2,3,... $. The choice
$\phi=\pi /2$ for the pump field then gives
\br
\til{a}(\tau '_n) &=&  b(0) e^{i(n+1)\pi} = e^{i\theta _n }b(0)\, ,
\label{changea}\\
\til{b}(\tau '_n) &=&  a(0) e^{in\pi} = e^{i\d _n }a(0) \, ,
\lb{changeb}
\er
thus identifying $\theta _n$ and $\d _n$. The {\em exchange time}, 
$T'= \pi /(2\l)$, is the fundamental period when operators are
exchanged, it is half the recurrence time when one sets $\O =0$,
$T=2T'$. Summarizing, when this resonance condition is introduced
into the problem the time-evolving operators $\til{a} (\tau)$ and
$\til{b} (\tau)$, will recur at times
multiples of $\pi/\l$ and exchange identity at odd multiples of
$\pi/(2\l)$ (except for a phase factor).

%
\section{Joint state vector in the RWC: recurrence and exchange
of identity}
%
Now we are proceeding to calculate the time evolution of the two-mode
{\it state vector\/}, with the goal
to determine the times at which the system recurs and the
times when the modes exchange their states, i.e., each one assuming
the state of the other at $\tau=0$.
We confine ourselves to a special class of initial
states, which can be described as a finite superposition of the
coherent states. In this case one can easily transform the expressions
for the time-dependent Heisenberg operators found above to the
expressions for the state vectors,
using the method of {\it characteristic functions}
\cite{mollow2,cahill1,cahill2}.

The symmetric form of the characteristic function is
\begin{eqnarray}
\chi _{S}(\eta ,\zeta ,t) &=&{\rm Tr}_{AB}\left[ \rho _{AB}(t)\;e%
^{\eta a^{\dagger }+\zeta b^{\dagger }-\eta ^{\ast }a-\zeta ^{\ast
}b}\right]  \nonumber \\
&=&{\rm Tr}_{AB}\left[ \rho _{AB}(0)\;e^{\eta a^{\dagger
}(t)+\zeta b^{\dagger }(t)-\eta ^{\ast }a(t)-\zeta ^{\ast }b(t)}\right] \, ,
\label{chisim1}
\end{eqnarray}
where the RHS of the first (second) line stands for the Schr\"odinger
(Heisenberg) picture, $\rho_{AB}$ being the density operator
for the two-mode system, $A+B$. Whenever the  field operators
$a(t)$ and  $b(t)$ (Heisenberg picture) depend linearly on
$a(0)$ and  $b(0)$ (Schr\"odinger picture) it becomes possible
to define new time-dependent functions
\be
\bar{\eta} \equiv \bar{\eta}(\eta ,\zeta ;t)\, , \qquad 
\bar{\zeta} \equiv \bar{\zeta}(\eta ,\zeta ;t)\, ,\lb{etazetabar}
\ee
and rewrite \rf{chisim1} as
\be
\chi _{S}(\eta ,\zeta ,t) ={\rm Tr}_{AB}\left[ \rho _{AB}(0)\;e%
^{\bar{\eta}a^{\dagger }+\bar{\zeta}b^{\dagger }-\bar{\eta}^{\ast
}a-\bar{\zeta}^{\ast }b}\right]
\equiv \chi _{S}(\bar{\eta},\bar{\zeta},0) \,. \lb{chisim2}
\ee
The normal form of the characteristic function can be found
through the relation
\begin{equation}
\chi _{N}(\eta ,\zeta ,t)=e^{\frac{1}{2}\left( \left|
\eta \right| ^{2}+\left| \zeta \right| ^{2}\right) }\chi _{S}(\eta ,\zeta
,t)\,,  \label{chinorm1}
\end{equation}
which can be written in terms of \rf{etazetabar} as
\begin{equation}
\chi _{N}(\eta ,\zeta ,t)=e^{\frac{1}{2}\left[ \left|
\eta \right| ^{2}+\left| \zeta \right| ^{2}-\left| \bar{\eta}\right|
^{2}-\left| \bar{\zeta}\right| ^{2}\right] }\chi _{N}(\bar{\eta},\bar{\zeta}%
,0),  \label{chinorm2}
\end{equation}
with
\begin{equation}
\chi _{N}(\bar{\eta},\bar{\zeta},0)=e^{\frac{1}{2}\left[
\left| \bar{\eta}\right| ^{2}+\left| \bar{\zeta}\right| ^{2}\right] }\chi
_{S}(\bar{\eta},\bar{\zeta},0)\,.  \label{chinorm3}
\end{equation}
Substituting Eqs. \rf{etazetabar} in the second line of Eq. \rf{chinorm1}
and after rearranging the terms we obtain
\be
\chi _{S}(\eta ,\zeta ,t) ={\rm Tr}_{AB}\left[ \rho _{AB}(0)\;e%
^{\bar{\eta}a^{\dagger }-\bar{\eta}^{\ast }a}e^{%
\bar{\zeta}b^{\dagger }-\bar{\zeta}^{\ast }b}\right]
=\chi _{S}(\bar{\eta},\bar{\zeta},0) \, ,  \label{chisim4}
\ee
where the dynamical evolution is present only in the parameters
\begin{eqnarray}
\bar{\eta} =  \eta \tilde{u}_{1}^{\ast}+
\zeta \til{v}_{2}^{\ast },  \label{etabar2} \\
\bar{\zeta} =  \eta \tilde{v}_{1}^{\ast}
+\zeta \tilde{u}_{2}^{\ast } \, ,  \label{zetabar2}
\end{eqnarray}
\be
\tilde{u}_1\equiv e^{-i\o_a t}u_1(\tau), \quad
\tilde{v}_1\equiv e^{-i\o_a t}v_1(\tau), \qquad
\tilde{u}_2\equiv e^{-i\o_b t}u_2(\tau), \quad
\tilde{v}_2\equiv e^{-i\o_b t}v_2(\tau).
\label{uvtil}
\ee
From the definitions \rf{etabar2} and \rf{zetabar2} one verifies that
\begin{equation}
\left| \bar{\eta}\right| ^{2}+\left| \bar{\zeta}\right| ^{2}=\left| \eta
\right| ^{2}+\left| \zeta \right| ^{2}\, ,  \label{igualdade1}
\end{equation}
then equation \rf{chinorm2} becomes
\begin{equation}
\chi _{N}(\eta ,\zeta ,t)=\chi _{N}(\bar{\eta},\bar{\zeta},0) \,.
\lb{chi=chi}
\end{equation}
If the initial joint density operator is factorised (absence of initial
correlations),
\begin{equation}
\rho _{AB}(0)=\rho _{A}(0)\ox \rho _{B}(0) \, , \lb{rho2}
\end{equation}
then the normal joint characteristic function (Fourier space)
factorises as
\begin{eqnarray}
\chi _{N}(\eta ,\zeta ,t) &=&{\rm Tr}_{A}\left[ \rho _{A}(0)\;e%
^{\bar{\eta}a^{\dagger }}e^{-\bar{\eta}^{\ast
}a}\right] \;{\rm Tr}_{B}\left[ \rho _{B}(0)\;e^{\bar{\zeta}%
b^{\dagger }}e^{-\bar{\zeta}^{\ast }b}\right]  \nonumber
\\
&=&\chi _{N}^{A}(\bar{\eta},0)\;\chi _{N}^{B}(\bar{\zeta},0) \,.
\label{chiAchiB}
\end{eqnarray}
Thus, contrarily to the density operator which correlates in the course
of its evolution, in the Fourier space the modes evolve in an apparent
uncorrelated fashion, however there is a hidden correlation which is
present in the parameters \rf{etabar2} and \rf{zetabar2}, since they depend
on the functions \rf{u} and \rf{v}.

We assume that the mode $A$ was prepared `initially' ($\tau=0$) in a
superposition of two coherent states,
\begin{equation}
\left| \Psi _{A} (t;\tau=0)\right\rangle =\frac{1}{N}\left( \left| \alpha
e^{-i\o _a t}\right\rangle +e^{i\Phi} \left| -\alpha e^{-i\o _a t}
\right\rangle \right) \, , \quad
N = \sqrt {2 \lpar 1+ \cos \Phi e^{-2|\alf |^2}  \rpar} \, .
\lb{gato}
\end{equation}
The special cases of the state (\ref{gato})
for $\Phi =0, \pi/2, \pi$  are known as {\em even
cat state} \cite{dodonov}, {\em Yurke-Stoler state} \cite{yurke}
and {\em odd cat state} \cite{dodonov}, respectively.
The `initial' density operator
is therefore $\rho _{A}(t;\tau=0)=\left| \psi _{A}(t;\tau=0)\rg \lg
\psi _{A}(t;\tau=0)\right| $.
The mode $B$ is assumed to be `initially' in the coherent state
$|\beta e^{-i\o _b t}\rg$ with $\rho _{B}(t;\tau=0) =
\left| \beta (t;\tau=0)\rg \lg \beta (t;\tau=0) \right|$. By `initial' we
do not mean $t=0$, but when the interaction is yet not turned on, the
slow motion begining at $\tau=0$ when $\l =0$ (not $t=0$).

Substituting $\rho _{A}(t;0)$ and $\rho _{B}(t;0)$ into the characteristic
function \rf{chiAchiB} one obtains
\be
\chi _{N}(\eta ,\zeta ,t) = \frac{1}{N^{2}}\left[ \;e^{%
\bar{\eta}\alpha ^{\ast }-\bar{\eta}^{\ast }\alpha }+e^{-%
\bar{\eta}\alpha ^{\ast }+\bar{\eta}^{\ast }\alpha }+e^{-2|\alf|^2}
\lpar e^{i\Phi} e^{
\bar{\eta}\alpha ^{\ast }+\bar{\eta}^{\ast }\alpha }+e^{-i\Phi}e^{-%
\bar{\eta}\alpha ^{\ast }-\bar{\eta}^{\ast }\alpha } \rpar \right]  
e^{\bar{\zeta}\beta ^{\ast }-\bar{\zeta}^{\ast} \beta } \, .
\nonumber
\ee
Using equations \rf{etabar2} and \rf{zetabar2} and
after some algebraic manipulation one gets
\begin{eqnarray}
\chi _{N}(\eta ,\zeta ,t) &=&\frac{1}{N^{2}} \left\{ \exp \;\left[
\eta  z_1^{*}- \eta^* z_1 +\zeta z_3^*  -\zeta ^{*} z_3  \right] 
+\exp \;\left[ -\eta z_2 ^* + \eta ^{*} z_2 -\zeta z_4^* +
\zeta ^{*} z_4 \right]  \right.
\nonumber \\
&+& \left. e^{-2\left| \alpha \right| ^{2}} \left(  e^{i\Phi} \exp \;\left[
\eta z_1^* +\eta ^{*} z_2 +\zeta z_3^* +\zeta^{*} z_4 \right]  
+ e^{-i\Phi} \exp \;\left[ -\eta z_2^* -\eta ^{*} z_1
-\zeta z_4^* -\zeta ^{*} z_3 \right] \right) \right\} \,
\label{chinorm4}
\end{eqnarray}
where
\be
z_1 = \tilde{u}_1 \alf + \tilde{v}_1 \beta  \, , \quad
z_2 = \tilde{u}_1 \alf - \tilde{v}_1 \beta  \, , \quad
z_3 = \tilde{v}_2 \alf + \tilde{u}_2 \beta  \, , \quad
z_4 = \tilde{v}_2 \alf - \tilde{u}_2 \beta  \, ,
\ee
thus the time-dependent new labels $z_i $ are the linear superpositions
of the initial ones, $\alf, \beta $, for the joint-state written in
the Schr\" odinger picture, where the multiplicative time-dependent phases
$e^{-i\o_i t}$ of the functions $\tilde{u}_i, \tilde{v}_i$ ($i=a,b$) in
the definition \rf{uvtil}
stand for the `fast' oscillation of the field whereas
the time dependence in $u_i(\tau)$ and $v_i(\tau)$ correspond to the
`slow' motion of the field configuration.
We can write the two-mode normal characteristic function in the
Schr\"odinger picture also as
\begin{equation}
\chi _{N}(\eta ,\zeta ,t)={\rm Tr}_{AB}\left[ \rho _{AB}(t)\;
e^{\eta a^{\dagger }}e^{-\eta ^{\ast }a}
e^{\zeta b^{\dagger }}e^{-\zeta ^{\ast }b}\right]
= \left\langle \Psi_{AB} (t)\right| \;e^{\eta a^{\dagger }}
e^{\zeta b^{\dagger }}e^{-\eta ^{\ast }a}e^{-\zeta ^{\ast }b}\left|
\Psi_{AB} (t)\right\rangle
\, .\label{chinorm5}
\end{equation}
For a superposition $ | \Psi_{AB} (t)\rg =e_{1} |\Psi _{1}(t)\rg
+e_{2} | \Psi _{2}(t)\rg $, a comparison between Eqs.
\rf{chinorm5} and \rf{chinorm4} permits to identify the joint statevector,
at any time $t$, showing an entanglement of the modes, 
\be
|\Psi_{AB}(t) \rg = \frac 1N \lpar |z_1 , z_3 \rg + e^{i\Phi}
|-z_2 , -z_4 \rg \rpar
\equiv \frac{1}{N}\left( |z_1 \rg _{A} \ox |z_3 \rg _{B}
+ e^{i\Phi} | - z_2 \rg _{A}\ox | -z_4 \rg _{B} \right) \, ,
\label{estadoAB}
\ee
where the $|\pm z_i \rg $ are coherent states.

The time-dependent functions  \rf{u} and \rf{v} calculated at
`slow' recurrence times $\tau_n = n\pi$, with $n=1,2,3,...$,
are
\begin{equation}
u_{1}(\tau_{n})=e^{in\pi(1+\chi) },\quad
u_{2}(\tau_{n})=e^{in\pi(1-\chi) }, \quad
v_{1}(\tau_{n})= v_{2}(\tau_{n})=0 \,.
\end{equation}
So the joint statevector \rf{estadoAB} is
\be
| \Psi_{AB} (t;\tau_{n})\rg = \frac 1N \lpar | \alpha_n(t)  \rg
+ e^{i\Phi} | -\alpha_n(t) \rg \rpar _{A} \ox
| \beta_n(t) \rg _{B} \, , \lb{estadorecor}
\ee
with
\be
\alf _n(t) \equiv \alf \exp\left[-i\o _a t +in\pi(1+\chi)\right], \quad
\beta _n(t) \equiv \beta \exp\left[-i\o _b t +in\pi(1-\chi)\right] \, .
\label{defanbn}
\ee
Thus $|\Psi_{AB} (t;\tau_{n}) \rg $ has the same {\it functional form}
as the initial state, although the positions of the centers of the peaks
are rotated in the complex planes $\alpha$ and $\beta$. However, exact
recurrence of the fields can happen for the
choice $\chi = m/n~ (m<n),~m=1,2,3,...~ {\rm and}~ n=m+1,m+2,...$, $m+n=$
even, thus one gets $\alf _n(t) = \alf e^{-i\o _a t}$, and $\beta _n(t)=
\beta  e^{-i\o _b t} $.

For $\O=\chi=0$, $\phi=\pi/2$ and at `slow' times $\tau_n^{\prime} =
(n - 1/2)\pi$ with $n=1,2,3,...$, the functions \rf{u} and \rf{v}
take the values
\begin{equation}
u_{1}(\tau '_n)=u_{2}(\tau '_n)= 0 \, , \quad
v_{1}(\tau '_n)=(-1)^{n+1} \, , \quad v_{2}(t'_n)= (-1)^{n} \, ,
\end{equation}
so, the joint statevector writes
\be
\left| \Psi_{AB} (t;\tau'_n)\right\rangle =
|(-1)^{n+1}\beta e^{-i\o _a t} \rg _{A}\ox \frac 1N
\left( |(-1)^n \alpha e^{-i\o _b t} \rg +
e^{i\Phi}| -(-1)^n \alpha e^{-i\o _b t} \rg \right) _{B}  \, .  \lb{Psi1}
\ee
Now the initial superposition state (mode $A$) and coherent state
(mode $B$) do an exchange of configuration at `slow' times
$\tau'_n=(n-1/2)\pi $, although the phases of the field variables are
not exactly the same as in the original state (at $\tau=0$).
For $n=2,4,6,...$ the state of mode $A$ is $|-\beta e^{-i\o _a t} \rg$,
thus the
field acquires an extra phase (it suffers a rotation by an angle $\pi$
in phase space) but the mode $B$ becomes {\em exactly} the
original superposition state of mode $A$. For $n=1,3,5,...$ the state of
mode $A$ is $|\beta e^{-i\o _a t}\rg$ and of mode $B$ is the superposition
$|-\alf e^{-i\o _b t}\rg +e^{i\Phi}|\alf e^{-i\o _b t}\rg$, where
$\alf$ changed sign, thus not
repeting the original superposition of mode $A$. However, for
$\Phi=0$ the state of mode $B$  becomes identical to the state of
mode $A$ at $\tau=0$ and vice versa; for $\Phi=\pi$ the only difference
is that state $B$ becomes equal to the initial superposition of mode
$A$ times a global phase (a minus sign). So, for $\Phi=0,\pi$ and at
times $t_n, n={\rm odd}$ an exchange of state configuration between the
two modes takes place, the coherent state goes to the superposition
state and vice-versa, this shows that (at $\tau _n$'s) there is a full
transfer of information among the modes, allowing an exchange of identity.
Moreover, it is worth observing that in \rf{Psi1}
$\alf $ and $\beta  $ are multiplied by the factors $e^{-i\o _b t}$
and $e^{-i\o _a t}$, respectively, thus the frequency of
the fast oscillating factor goes along with the mode and not with the
label of the initial state, so it is this factor who gives a signature
to the mode and not the `shape' of the field.

We remind that the exchange of identity
between the modes is possible only for $\O =0$, which means that the
frequency $\nu$ of the pump field must be `resonant' with the
difference of frequencies  $\o_ a - \o _b$ (if $\o _a > \o _b$,
otherwise one should choose $\nu = \o _b - \o _a$). At `slow' times
$\tau_n^{\prime} = \pi(n -1/2) $ the states are
maximally correlated, attaining the maximal entanglement.
%
\section{Decoherence and entropy}
%
In order to analyse the decoherence of a single mode, for instance
mode $A$, we have to follow the time evolution of its state 
independently of that for mode $B$. The joint density operator is
\be
\rho _{AB}(t) = \frac{1}{N^{2}} \left( | z_1, z_3 \rg \lg z_1, z_3| 
+| -z_2, - z_4 \rg \lg - z_2, - z_4 | + e^{i\Phi} | z_1, z_3 \rg
\lg - z_2,- z_4 | + e^{-i\Phi}| - z_2, - z_4 \rg \lg z_1 , z_3 | \right)
\, , \label{rototal}
\ee
and the calculation of $\rho _{A}={\rm Tr}_{B}\left[ \rho _{AB}\right]$
gives the reduced density operator (hereafter we set $\O=0$)
\br
\rho _{A}(t) &=&\frac{1}{N^{2}}\left\{ \left| z_1
\right\rangle \left\langle z_1 \right| +\left| - z_2
\right\rangle \left\langle - z_2  \right| \right. \nn \\ 
&& \left. +
\exp{\left[ -\left| \alpha \right| ^{2}
\left( 1-\cos \left(2\lambda t\right) \right) \right]}
\left[ e^{-i\Phi} e^{2i{\rm Im}\lpar u_2 v^*_2 \alf ^* \beta \rpar}
\left| z_1 \right\rangle
\left\langle - z_2 \right| + H.c. \right] \right\} \, .  \lb{roA}
\er
The first exponential factor is responsible
for the decoherence (reduction of the nondiagonal terms in \rf{roA}).
The decoherence process begins altogether with the evolution of the
two modes, and the larger is $|\alf|$ the more pronounced will be
the decoherence, attaining its climax at $T_d = \pi/(2\l)$
(the exponential factor becomes $\exp(-2|\alf|^2)$), then it begins
its way back, the state of mode $A$ recohers,
returning to the initial state at $T_r = \pi/\l$. The periodicity
of decoherence and recoherence occurs because the `environment' of
mode $A$ is constituted by a single mode ($B$) \cite{raimond}; for a
multimode environment the recoherence time would be quite larger and for
an infinite number of modes the decoherence becomes irreversible
\cite{caldeira,zurek}.

The linear entropy
\begin{equation}
S_{A}(t)={\rm Tr}_{A}\left[ \rho _{A}(t)-\rho _{A}^{2}(t) \right] \, ,
\lb{entropia}
\end{equation}
is well suited for the analysis of the decoherence; from the norm
conservation, ${\rm Tr}_{A}\left[ \rho _{A}(t)\right] =1$,
\rf{entropia} becomes \mbox{$S_{A}(t)=1 - {\rm Tr}_{A}\rho _{A}^{2}(t)$}.
For a pure state  $\rho _{A}^{2}(t)=\rho _{A}(t)$,
${\rm Tr}_{A}\rho _{A}^{2}(t) =1$, thus $ S_{A}(t)=0 $.
For a statistical mixture
\mbox{$ 0 \leq {\rm Tr}_{A}\left[ \rho _{A}^{2}(t)\right] \leq 1 $}, then
$0 \leq S_{A}(t) \leq 1$. For \rf{roA} the entropy writes as
\br
S_A(t) = 1 - \frac {2}{N^4} && \lbk  1 +
e^{-2|\alf|^2 (1+\cos ( 2\l t))} + e^{-2|\alf|^2 (1-\cos (2\l t))}
+ 4 \cos \Phi ~ e^{-2|\alf|^2} \right. \nn \\ 
&& \left.
+  e^{-4|\alf|^2}
\cos \lbk 2\Phi ~ + 2 \sin \lpar 2\l t \rpar {\rm Im}(\alf \beta^*)
\rbk \rbk \, , \lb{entropyA} 
\er
and it is plotted in Fig.~1 for Im$(\alf \beta ^*) = 0$, $\Phi =0,
\pi/2, \pi$ and $|\alf|^2 =1$, where one can see the decoherence-recoherence
cycles. Initially the entropy is minimum ($S_A=0$), when the states of modes
are uncorrelated, then the initial pure state decoheres
and becomes a mixed state (the ascending curve), the nondiagonal terms
of the density operator attain their lowest value (the maxima of the
curve), then they increase back, recohering (the descending curve) and
the reduced density operator for $A$ becomes a pure state
when again $S_A =0$ at $\l t=\pi/2$, and the joint state disentangles
(in particular, when recurrence or exchange of identity of the
states takes place). This cycle repeats with period $\pi/2\l$.
Since the two mode are symmetrical the entropy of mode $B$ is the same as
\rf{entropyA}, $S_B(t) = S_A(t)$, thus during the interaction the
information flows at the same rate in the two `directions',
$A \Longleftrightarrow B$. In Fig. 2 we show the entropy with the
same settings as in Fig. 1, except that here $|\alf|^2 =5$; the three
`cat' states have nearly the same entropy and the duration of the
entangled state is larger (the plateau at the maxima) compared
to curves in Fig.~1.

The mean energy of two modes (for the state \rf{rototal})
are given by
\begin{eqnarray}
\lg E _A(t)\rg &=&{\rm Tr}_{AB}\left[ \rho _{AB}(t) \hbar \o _a
a^{\dagger }a \right] = \frac{2\hbar \o _a}{N^2} \left[ |\alpha | ^2
\cos ^2(\l t) \lpar 1 -\cos \Phi ~ e^{-2 |\alpha | ^2} \rpar \right. \nn \\
&& + \left. |\beta | ^2 \sin^{2}(\l t) \lpar 1 +
\cos \Phi ~e^{-2|\alpha | ^2} \rpar
+ {\rm Im} \lpar \alf \beta ^* \rpar \sin \lpar 2\l t \rpar \sin \Phi ~
e^{-2 |\alf|^2} \right] \, , \lb{enerA}
\end{eqnarray}
\begin{eqnarray}
\left\langle E_{B}(t)\right\rangle &=&  {\rm Tr}_{AB}
\left[ \rho _{AB}(t)\hbar \o _b b^{\dagger }b\right]  
=\frac{ 2\hbar \o _{b} }{ N^2 }\left[ | \alpha | ^{2}
\sin ^{2}(\l t)\left( 1 - \cos \Phi ~ e^{-2| \alpha | ^2} \right) \right.
\nn \\
&& + \left. |\beta | ^2 \cos ^{2}(\l t)\left( 1 + \cos \Phi ~
e^{-2|\alpha | ^2} \right)
- {\rm Im} \lpar \alf \beta ^* \rpar \sin \lpar 2\l t \rpar \sin \Phi ~
e^{-2 |\alf|^2} \right] \, , \lb{enerB}
\end{eqnarray}
both mean energies oscillate in time but out of phase by $\pi$, 
their sum oscillates in time too since $\o _a \neq \o _b$.
For the interaction \rf{girante} the total number of photons of
the two modes, $n_A+n_B = a^\dag a + b^\dag b$, is a conserved
quantity, however, the number of photons of each mode is not,
\begin{equation}
\left[ H,n_A+n_B\right] =0 \, , \quad \left[ H,n_A\right] \neq 0 \, ,
\quad \left[ H,n_B\right] \neq 0 \, . 
\end{equation}
The expectation value of $n_A + n_B $ in \rf{rototal} is given by
\begin{equation}
\bar{n}_{A}+\bar{n}_{B}=\frac{\left\langle E_{A}(t)\right\rangle }{\hbar
\omega _{a}}+\frac{\left\langle E_{B}(t)\right\rangle }{\hbar \omega _{b}}=
\frac{2}{N^{2}}\left[ \left| \alpha \right| ^{2}\left( 1 -\cos \Phi ~ e%
^{-2\left| \alpha \right| ^{2}}\right) +\left| \beta \right|
^{2}\left( 1 + \cos \Phi ~ e^{-2\left| \alpha \right| ^{2}}\right)
\right] \, ,
\end{equation}
and is a constant of motion.

%
\section{Information exchange without intermode energy transfer}
%
In general, decoherence of a subsystem is presented altoghether with
energy flow to the ``other" degrees of freedom (the environment)
\cite{caldeira}, thus transference of information between two interacting
subsystems goes along with a transference of energy. For instance,
the processes of decoherence and damping of a system coupled to a
thermal reservoir is interpreted as a leak of information from the
system to the environment altogether with a loss of energy,
the dynamical evolution terminating with the thermalization of the
system \cite{zurek}. Now we shall verify that for some particular
relation between $\alf$ and $\beta$, the transference of information
between the modes may happen without energy transfer.

Looking at the RHS of Eqs.\rf{enerA} and \rf{enerB} one notes that if
the modes are prepared such that ${\rm Im} (\alf \beta^*) = 0$,
i.e. $\alf$ and $\beta$ have the same phase, and the intensities
are related by the relation
\be
|\beta|^2 = |\alf|^2 \frac{1 - \cos \Phi ~ e^{-2|\alf|^2}}
{1 + \cos \Phi ~ e^{-2|\alf|^2}} \, ,
\ee
each mean energy becomes constant in time. In particular, 
for $\Phi = 0$, $|\beta|^2 = |\alf|^2 \tanh|\alf|^2$, when
\begin{equation}
\left\langle E_{A}(t)\right\rangle =  \hbar \omega _{a}\left|
\alpha \right| ^{2}\tanh(|\alpha |^{2}) \, , \qquad
\left\langle E_{B}(t)\right\rangle = \hbar \omega _{b}\left|
\alpha \right| ^{2}\tanh(|\alpha |^{2}) \, ,
\end{equation}
and for $\Phi = \pi$, $|\beta|^2 = |\alf|^2 \coth|\alf|^2$, when
\be
\left\langle E_{A}(t)\right\rangle =  \hbar \omega _{a}\left|
\alpha \right| ^{2}\coth(|\alpha |^{2}) \, , \qquad
\left\langle E_{B}(t)\right\rangle = \hbar \omega _{b}\left|
\alpha \right| ^{2}\coth(|\alpha |^{2}) \,.
\ee
So, optimal transference of information between modes (in the sense
of exchange of states configuration) without energy transfer occurs for a
particular relation between the field amplitudes, namely, for
the modes possessing the same mean number of photons, with conservation of
mean energy for each mode and the ratio between mean energies,
\begin{equation}
\frac{\left\langle E_{B}(t)\right\rg}{\left\langle E_{A}(t)\right\rg}
 = \frac{\omega _{b}}{\omega _{a}}
\end{equation}
depending only on the frequencies ratio.

However, the variance of the mean number of photons of each mode,
$n_A$ and $n_B$ is not null neither constant, but it oscillates in time,
for instance,
\br
V_{\Phi=0, \pi}(n_A) &\equiv& \lg n_A ^2 \rg -\lg n_A \rg ^2 =
|\alf| ^4 \lbk \cos ^4 (\l t) - \lpar 1- \sin ^4 (\l t) \rpar
\lpar \frac{1\mp e^{-2|\alf|^2}}{1\pm e^{-2|\alf|^2}}\rpar^2 
\right] \nn \\
&& + |\alf|^2 \cos ^2 (\l t)\lpar \frac{1\mp e^{-2|\alf|^2}}{1\pm
e^{-2|\alf|^2}}
\rpar \lbk 1 + 4 |\alf|^2 \lpar \frac{1\mp e^{-2|\alf|^2}}
{1\pm e^{-2|\alf|^2}}\rpar \sin ^2 (\l t) \rbk  \, ,  \lb{varA}
\er
where upper (lower) sign stands for $\Phi=0 (\pi)$, and for mode
$B$, one obtains an expression similar to \rf{varA}, however oscillating
out of phase. So, when there is no transference of energy 
between the modes the transference of information induces fluctuations
in the number of photons.
Therefore we can assert that if the modes have an equal mean number
of photons the information transfer occurs without energy transfer,
transference of energy taking place only when the number of photons is
unequal, however a fluctuation in their number is verified.
Since we assumed that the fields have the same phase,
Im$(\alf \beta ^*) = 0$, the expression for the
entropy \rf{entropyA} is the same whether the mean energies of the
modes are conserved or not,
so, the linear entropy \rf{entropia} measures the flow of
information from and to mode $A$, independently of its mean energy being
conserved or not. So, in this enlightning example we have seen that
decoherence is a process of pure information transfer 
without any need of transference of energy.

The time-dependent entropy \rf{entropyA} becomes equal to zero
at times when the joint state \rf{rototal} disentangles, however the
disentanglement means that either the joint state \rf{estadoAB} returned
to its initial state or an exchange of state took place. As we
wish to measure how much far is \rf{estadoAB} from an
exchange of state we define the {\em state-exchange functional}
\begin{equation}
{\mathcal E}(\tau)\equiv \frac{{\rm Tr}_{AB}\lbk \rho _{AB}(t;\tau)
\rho_{BA}(t;0)\rbk} {\lbk {\rm Tr}_{AB} ~\rho_{BA}(t;0) \rbk ^2} =
\frac{\left| \left\langle \Psi _{BA}(t;0)\right.
\left| \Psi _{AB}(t;\tau)\right\rangle \right| ^{2}}{\left| \left\langle \Psi
_{BA}(t;0)\right. \left| \Psi _{BA}(t;0)\right\rangle \right| ^{2}}\, ,
\lb{troca}
\end{equation}
(the second equality standing for pure states)
where $|\Psi_{AB} (t;0) \rg \equiv | \phi (t;0) \rg _{A} \ox |\chi(t;0)
\rg _{B}$, is the `initial' (remind that by initial, $\tau =0$, we mean
just before the interaction is turned on) joint state and
$|\Psi_{BA} (t;0) \rg \equiv | \chi (t;0) \rg _{A} \ox |\phi(t;0) \rg _{B}$
is the exchanged state and the denominator is introduced just to
normalize the functional to $1$, thus $0 \leq {\mathcal E}(\tau) \leq 1$.
The functional \rf{troca} becomes $1$ each time an exchange of states
occurs and if $|\phi(t;0) \rg $ and $|\chi (t;0) \rg $ are orthogonal
\rf{troca} becomes zero whenever the state $|\Psi _{AB}(t;\tau)\rg$ recurs
to the initial state. For $\O =0$ ${\mathcal E}(\tau)$ becomes
\br
{\mathcal E}(\tau) = &&\frac{2 e^{-2 \lbk |\alf|^2 +|\beta|^2
\lpar 1 - \sin \tau \rpar \rbk }}{N^4}
\lbr  e^{2\cos \tau \Re (\alf \beta ^* )} \cosh \lpar 2 \Re (\alf z_3^*)
\rpar + e^{-2\cos \tau \Re (\alf \beta ^* )} \cosh \lpar 2 \Re
(\alf z_4^*) \rpar + e^{-2 \sin \tau |\alf|^2} \right. \nn \\
&&\left. + 4 \cosh \lbk 2 \cos \tau \Re (\alf \beta ^*) \rbk
\cos \lbk \Phi - 2 \cos \tau \Im ( \alf \beta ^*) \rbk
+ e^{2 \sin \tau |\alf|^2} \cos \lbk 2 \lpar \Phi - 2 \cos \tau
\Im ( \alf \beta ^*) \rpar \rbk \rbr \, , \lb{troca2}
\er
and it is immediate to verify that at times $\tau _n =n\pi/2,~n=1,5,9,...$,
${\mathcal E}(\tau _n)$ assumes the same value,
\be
{\mathcal E} \lpar {\pi \over 2} \rpar = \lpar \frac{\cos \Phi +
e^{-2|\alf |^2}}{1 + \cos \Phi ~e^{-2|\alf |^2}} \rpar^2 \, ,  \lb{troca3}
\ee
the fact that it is independent of $\beta$ means that mode $B$
transfered exactly its original configuration (coherent state) to mode $A$,
but simultaneously mode $A$ cannot transfer its original `cat' state form
to mode $B$ for any value of $\Phi$, but only for the special cases of
even and odd `cat' states, $\Phi=0,\pi$, when ${\mathcal E}(\pi /2) =1$.
For $|\alf|^2 \gg 1$, ${\mathcal E}(\pi /2) = \cos^2 \Phi $, becoming
independent of $|\alf|$. The Yurke-Stoler state ($\Phi = \pi/2 $)
could not participate in the state-exchange scheme. 

At times $\tau _n = n\pi /2,~n=3,7,11,...$, ${\mathcal E}(\tau_n)$ assumes
always the same value,
\be
{\mathcal E}\lpar{3\pi \over 2}\rpar = e^{-4|\beta|^2} \, , \lb{troca4}
\ee
being independent of $\alf$ and $\Phi$, thus the mode $A$ transfered
information to mode $B$ in order to reproduce the original state of $A$,
but the inverse did not occured, the mode $A$ was unable to reproduce
exactly the state of mode $B$; an exact exchange of identity only
happens if the original state of mode $B$ is the vacuum state,
$\beta =0$, independently of the value the phase $\Phi$ can take.
In Fig. 3  we plotted \rf{troca} versus time for $\Phi =0$ and
$|\alf|^2 =5, |\beta|^2 = |\alf|^2 \tanh (|\alf|^2)$, that
illustrates our comments.
%
\section{Summary and conclusions}
%
We considered the problem of the coupling between two monochromatic modes
(treated quantically) pumped by another (classical) monochromatic field,
into a nonlinear cristal. For the RWC we
determined the periods of {\em recurrence} (when modes $A$ and $B$
return to their initial joint state) and of {\em state exchange}, when
information between the modes is exchanged, becoming possible that
mode $A$ assumes the initial state of mode $B$ and vice-versa.
We atributed to modes $A$ and $B$ the generic `cat state'
and the coherent state, respectively, as initial joint state; we
analysed its evolution during the interaction, verifying that an exact
interchange between the states at certain periodic times occurs only
when the `cat state' is even or odd, or, the interchange occurs for
any kind of `cat state' only if mode $B$ is initially in the vacuum
state. We introduced a time-dependent state-exchange functional, which
showed to be quite helpfull for measuring the degree of {\em exchangeness}
of the state as times go on. We calculated the reduced entropy of
mode $A$ and the linear entropy associated to it and analysed the periodic
decoherence and recoherence processes, and identified the maximum
entanglement between the modes for the maximum of the entropy and
disentanglement (at recurrence or state-exchange times) when the entropy
becomes zero.

As the mean energy of each of the modes oscillates in time, 
their sum remaining constant for arbitrary number of photons in each
mode, we verified that when the modes have the same mean number of photons
the mean energies become constant in time, although the joint state
goes on in its evolution, therefore intermode state-exchange (exchange
of information) is possible without transference of energy, although
the variance of the number of photons oscillates in time. We verified
that the expression for the entropy related to the reduced density
operator of mode $A$ does not change whether the mean energies are constant
or not, thus it becomes clear the entropy is a measure of the flux
of information, loss (decoherence) or gain (recoherence), without
discerning on the direction of the flux of energy. In sum, by going  back
to an `old problem' we analysed it from a different point of view from
the other papers in the literature, unveiling interesting new features
in the realm of pure information transfer between systems. As a last
remark, we are considering the dynamics of the single mode phase operator
\cite{pegg} which may lead to a deeper insight on the interaction
process between the quantum fields; the calculations are in progress and
results will be presented in a forthcoming publication.

%
\acknowledgments{MCO thanks FAPESP for total financial support and SSM
thanks CNPq and FINEP for partial financial support}
%
%

%
%
%
\begin{figure}[tbp]
\caption{The linear entropy of mode $A$ versus time in units of the
parameter $\l$ and $|\alf|^2=1$. The solid, dashed and dot-dashed lines
correspond to the even ($\Phi=0$), Yurke-Stoler ($\Phi = \pi/2$) and
odd ($\Phi = \pi$) `cat' states.}
\label{fig1}
\end{figure}
\begin{figure}[tbp]
\caption{The same settings as Fig.~1, except that $|\alf|^2 =5$. The
three curves of Fig.~1 are practically coincident.}
\label{fig2}
\end{figure}
\begin{figure}[tbp]
\caption{The state-exchange functional versus time in units of $\l$,
showing a coincident curve for even and odd `cat' states ($\Phi=0,\pi$)
and attaining value 1 for $\l t = \pi/2, 5\pi/2, 9\pi/2,...$~. }
\label{fig3}
\end{figure}
\begin{figure}[tbp]
\caption{The same as Fig.~3 but for Yurke-Stoler state $\Phi = \pi/2$,
the state-exchange functional never attains value 1.}
\label{fig4}
\end{figure}
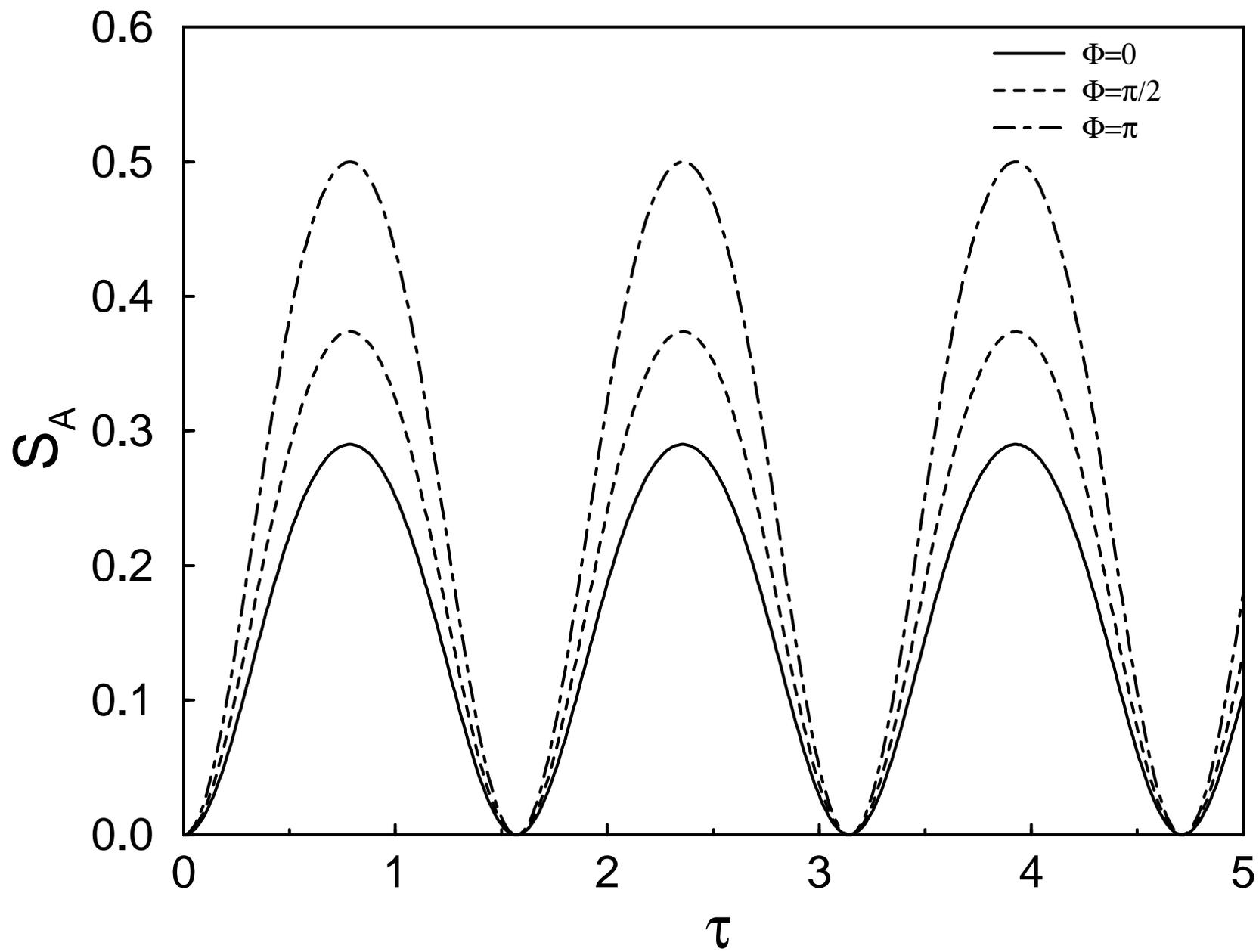
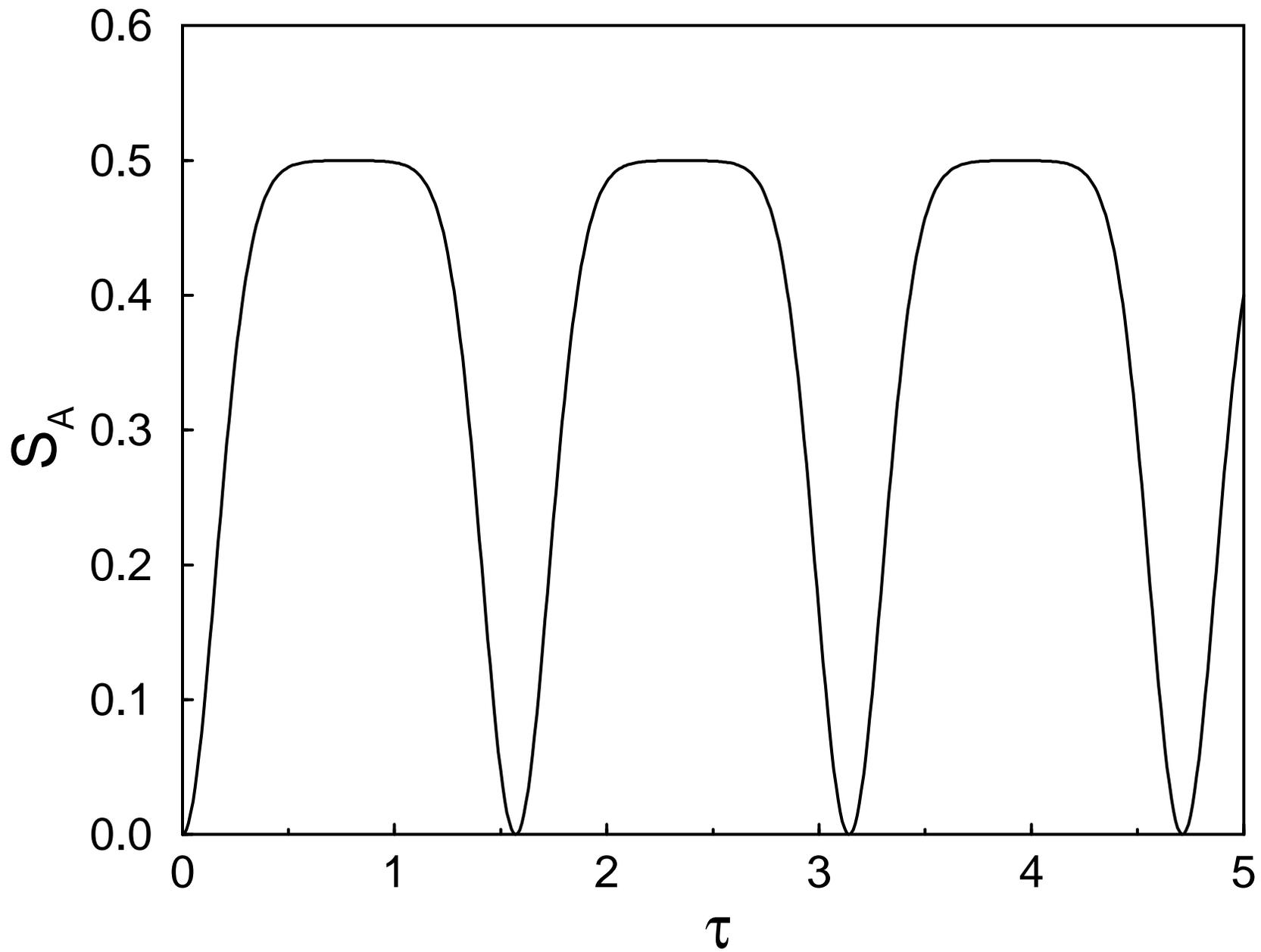
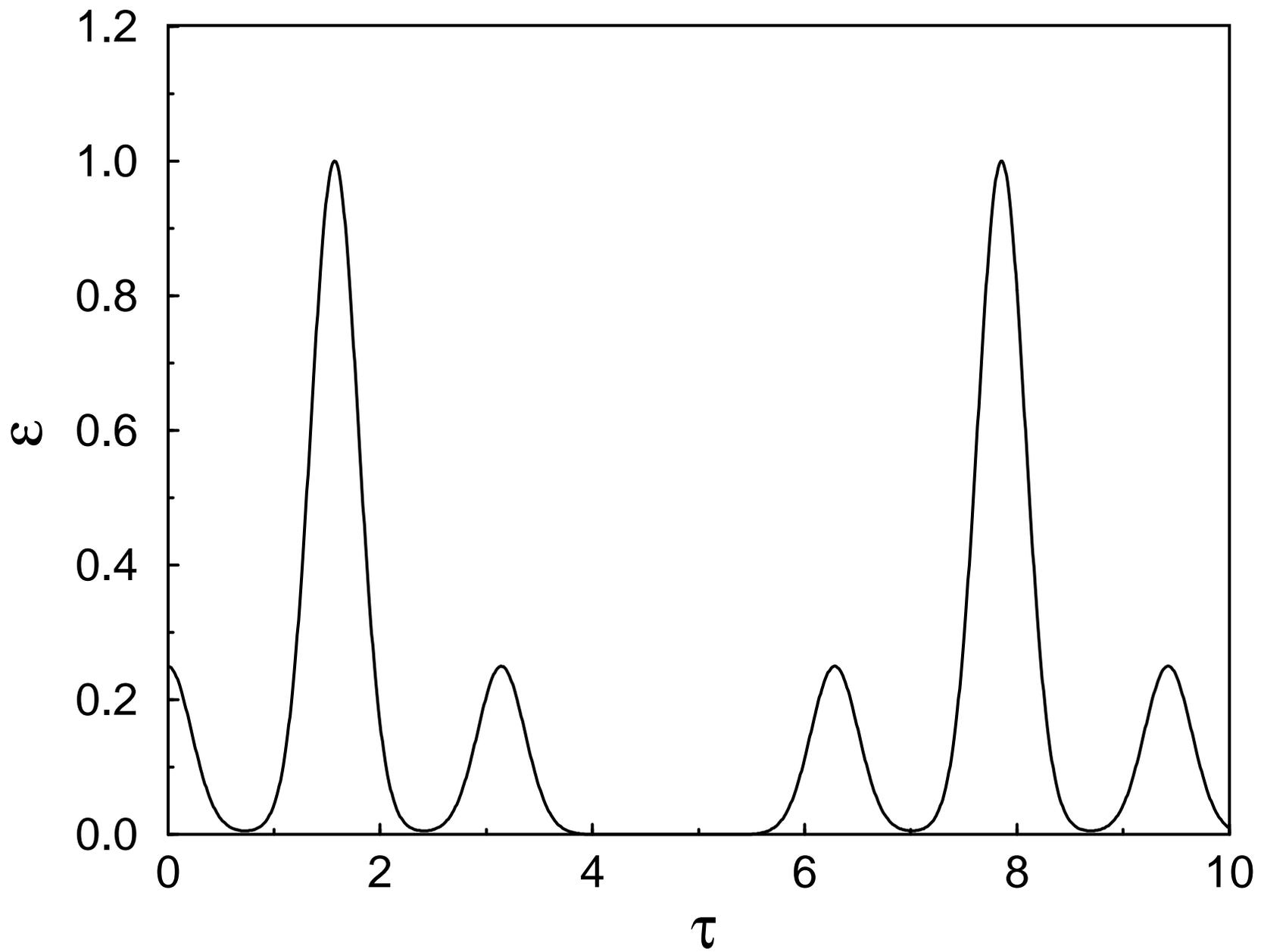
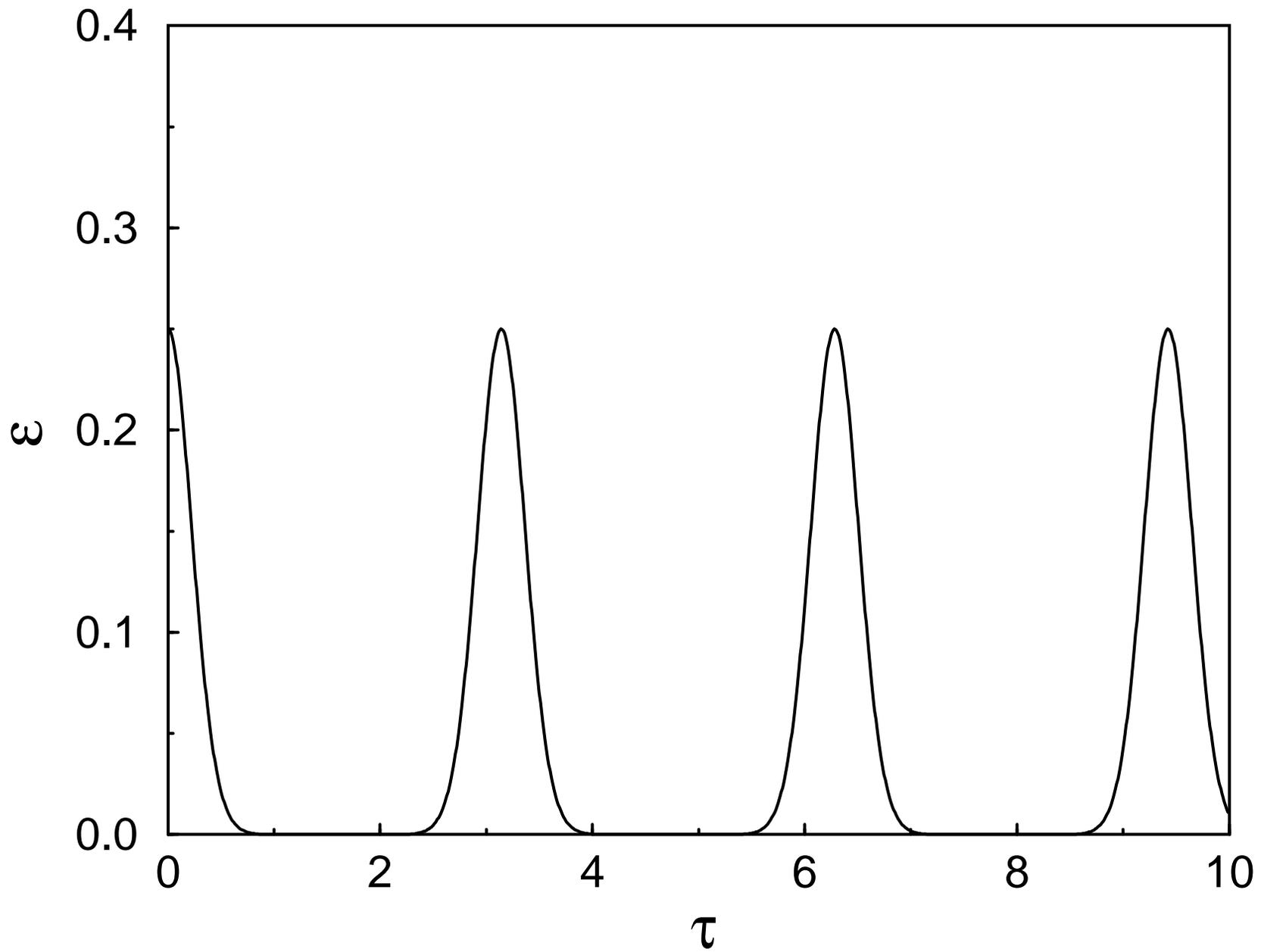
\end{document}